\documentclass{article}
\usepackage{graphicx}
\begin{document}

\title{\bf{Higgs Mass from g-2 in Anomaly
Mediated Supersymmetry Breaking Models}}

\author{Jun Tabei\footnote{jun@hep.phys.waseda.ac.jp} 
and Hiroshi Hotta\footnote{hotta@hep.phys.waseda.ac.jp} \\
Department of Physics, Waseda University, Tokyo 169, Japan }

\maketitle


\begin{abstract}
We estimate the upper bound of the Higgs mass 
and allowed parameter region 
in the anomaly mediated supersymmetry breaking models. 
There has been a difficulty that the parameter region 
cannot be determined precisely in almost SUSY models. 
However, we succeed to give strict constraints 
on the parameters from the recent results of the precision 
measurements of the muon g-2 (anomalous magnetic dipole moment). 
Especially, the upper bound of the Higgs mass is obtained 
as a function of only $\tan{\beta}$, and is estimated less than 
about 123 (GeV) for all the range of $\tan{\beta}<60$ . 
\end{abstract}

\section{Introduction}

The supersymmetry (SUSY) is one of the most 
attractive extension beyond the standard model (SM). 
It is important to detect the signals of SUSY not only in 
the high-energy region but also in low-energy experiments. 

In the high-energy physics, finding Higgs bosons 
is one of the main objectives for both of the SM and SUSY. 
As a common feature of SUSY 
models \cite{Espinosa,HEINEMEYER}, 
the Higgs boson mass is smaller than one in the SM. 
Thus, the search for the light neutral Higgs boson is 
regarded as a decisive test to give an evidence for SUSY 
that can be performed at present or in the next generation 
of high-energy colliders. 
The SM Higgs\footnote{the SUSY Higgs bosons are described 
as merely "Higgs" denoted by $m_{h}$,  
in contrast to the Higgs boson in the standard model $m^{SM}_{h}$ 
is expressed as "the SM Higgs" throughout this paper. } mass 
is currently restricted by LEP \cite{LEPSM,Particle} 
as follows: 

\begin{equation}
m^{SM}_{h} > 114.4 \ \mbox{(GeV)} .
\end{equation}
\noindent
On the other hand, the MSSM lightest neutral Higgs is constrained 
as \cite{Particle,LEPMSSM}: 

\begin{equation}
m_{h} > 90.1 \ \mbox{(GeV)} .
\end{equation}
\noindent
$m_{h}$ has been expected less than about 130 (GeV) 
in many SUSY models 
theoretically \cite{Espinosa,HEINEMEYER,DEDES}. 
However, some of the parameters in such SUSY models have been 
assumed loosely in almost analyses. 
In this paper, we show the allowed region of the parameters 
in more reliable way to be consistent 
with the low-energy precise physics such as 
the measurements of the muon anomalous magnetic dipole moment. 

The anomalous magnetic dipole moment (g-2) of the muon, 
defined as $a_{\mu}=(g_{\mu}-2)/2$ for convention, 
is one of the most precisely measured quantities 
in the low-energy particle physics. 
The world sum average, dominated by the recent measurements of 
the Muon (g-2) Collaboration 
at Brookhaven National Laboratory \cite{Bennett}, reads, 

\begin{equation}
a_{\mu}^{exp}=11659203 \pm 7 \times 10^{-10} \ . 
\end{equation}
\noindent
The difference between the experiment 
and the theoretical prediction in the SM of the muon g-2 
is currently \cite{Harbach}, 

\begin{equation}
a_{\mu}^{exp}-a_{\mu}^{SM}(e^{+}e^{-})=
(35.5 \pm 11.7) \times 10^{-10} \ , 
\label{eq:aepm}
\end{equation}
\begin{equation}
a_{\mu}^{exp}-a_{\mu}^{SM}(\tau)=
(10.3 \pm 10.7)  \times 10^{-10} \ , 
\label{eq:atau}
\end{equation}
\noindent
where, $e^{+}e^{-}$ and $\tau$ mean 
the kinds of the hadronic contributions to the muon g-2 
based on $e^{+}e^{-}$ and $\tau$ cross sections, respectively. 
This inconsistency between $e^{+}e^{-}$ and $\tau$-based values 
stems from their different cross sections \cite{Akhmetshin}. 
As Eqs.(\ref{eq:aepm}) and (\ref{eq:atau}) imply, 
the muon g-2 in the SM is slightly but significantly lower 
than the experimental results \cite{Harbach}. 
Thus, we think of all the $10 \times 10^{-10}$ scale difference 
of $a_{\mu}$ between the experimental results of 
the muon g-2 and its theoretical prediction in the SM, 

\begin{equation}
a_{\mu}^{exp}-a_{\mu}^{SM}=a_{\mu}^{SUSY} \ , 
\label{eq:adiff}
\end{equation}
\noindent
as the contributions of SUSY effects, in this paper. 

As a realistic SUSY model, 
the anomaly mediated supersymmetry breaking (AMSB) 
scenario \cite{Randall} is adopted in this paper. This scenario is 
a mechanism of SUSY breaking via the super-Weyl anomaly. 
The AMSB models have four important parameters 
$\{$ $m_{3/2}$, $m_{0}$, $\tan{\beta}$, sgn($\mu$) $\}$, 
where, $m_{3/2}$ stands for the gravitino mass corresponding to 
the vacuum expectation value (VEV) of the auxiliary field 
for breaking SUSY at near the Planck scale. 
$m_{0}$ denotes the bulk mass 
and introduced to avoid the sleptons being tachyonic. 
$\tan{\beta}$ is defined as the ratio between VEVs 
of the two Higgs doublets as usual. 
sgn($\mu$) means the sign 
of the Higgs mixing parameter. 
For more detail of AMSB models, see \cite{Randall}. 

The main purpose of this paper is to estimate the upper bound of 
the Higgs mass by restricting the allowed parameter region 
in AMSB models from precisely measured muon g-2. 

\section{Mass Spectra} 
In AMSB models, the soft SUSY breaking masses 
in the diagonal parts of the mass matrices of each scalar particles 
are given as follows \cite{Randall,Gherghtta,Huitu} : 

\begin{eqnarray}
m^{2}_{\tilde{Q}_{i}}
&=&c_{Q}m^{2}_{0}+\left(-\frac{11}{50}\alpha^{2}_{1}
-\frac{3}{2}\alpha^{2}_{2}+8\alpha^{2}_{3}
+\beta_{t}\delta_{i,3}+\beta_{b}\delta_{i,3}\right)
\frac{m^{2}_{3/2}}{16\pi ^{2}} \qquad 
\mbox{(L-squarks)} \ , \qquad 
\label{eq:m2q} 
\\
m^{2}_{\tilde{U}_{i}}
&=&c_{U}m^{2}_{0}+\left(-\frac{88}{25}\alpha^{2}_{1}
+8\alpha^{2}_{3}
+2\beta_{t}\delta_{i,3}\right)
\frac{m^{2}_{3/2}}{16\pi ^{2}} \qquad
\mbox{(u-type R-squarks)} \ , 
\label{eq:m2u} 
\\
m^{2}_{\tilde{D}_{i}}
&=&c_{D}m^{2}_{0}+\left(-\frac{22}{25}\alpha^{2}_{1}
+8\alpha^{2}_{3}
+2\beta_{b}\delta_{i,3}\right)
\frac{m^{2}_{3/2}}{16\pi ^{2}} \qquad
\mbox{(d-type R-squarks)} \ , 
\label{eq:m2d} 
\\
m^{2}_{\tilde{L}_{i}}
&=&c_{L}m^{2}_{0}+\left(-\frac{99}{50}\alpha^{2}_{1}
-\frac{3}{2}\alpha^{2}_{2}
+\beta_{\tau}\delta_{i,3}\right)
\frac{m^{2}_{3/2}}{16\pi ^{2}} \qquad
\mbox{(L-sleptons)} \ , 
\label{eq:m2l} 
\\
m^{2}_{\tilde{E}_{i}}
&=&c_{E}m^{2}_{0}+\left(-\frac{198}{25}\alpha^{2}_{1}
+2\beta_{\tau}\delta_{i,3}\right)
\frac{m^{2}_{3/2}}{16\pi ^{2}} \qquad
\mbox{(R-sleptons)} \ , 
\label{eq:m2e} 
\\
m^{2}_{h_{u}}
&=&c_{h_{u}}m^{2}_{0}+\left(-\frac{99}{50}\alpha^{2}_{1}
-\frac{3}{2}\alpha^{2}_{2}
+3\beta_{t}\right)
\frac{m^{2}_{3/2}}{16\pi ^{2}} \qquad
\mbox{(u-type Higgs)} \ , 
\label{eq:m2hu} 
\\
m^{2}_{h_{d}}
&=&c_{h_{d}}m^{2}_{0}+\left(-\frac{99}{50}g^{4}_{1}
-\frac{3}{2}g^{4}_{2}
+3\beta_{b}+\beta_{\tau}\right)
\frac{m^{2}_{3/2}}{16\pi ^{2}} \qquad
\mbox{(d-type Higgs)} \ , 
\label{eq:m2hd} 
\end{eqnarray}

with 

\begin{eqnarray}
\beta_{t}
&=&
\alpha_{t}\left(-\frac{13}{15}\alpha_{1}
-3\alpha_{2}-\frac{16}{3}\alpha_{3}+6\alpha_{t}+\alpha_{b}
\right) \ , 
\label{eq:bt}
\\
\beta_{b}
&=&
\alpha_{b}\left(-\frac{7}{15}\alpha_{1}
-3\alpha_{2}-\frac{16}{3}\alpha_{3}+\alpha_{t}+6\alpha_{b}
+\alpha_{\tau}
\right) \ , 
\label{eq:bb}
\\
\beta_{\tau}
&=&
\alpha_{\tau}\left(-\frac{9}{5}\alpha_{1}
-3\alpha_{2}+3\alpha_{b}+4\alpha_{\tau}
\right) \ , 
\label{eq:btau}
\end{eqnarray}
\noindent
where, the index $i$ indicates the generation. 
$\alpha_{n}=g^{2}_{n}/4\pi$ ($n$=1, 2, and 3 ) denotes 
the gauge coupling constant, and 
$\alpha_{f}=y^{2}_{f}/4\pi$ ($f$=$t$, $b$, and $\tau$) stands for 
the Yukawa coupling constant of top, bottom, and tau, respectively. 
$\delta_{i,j}$ is the Kronecker's delta. 
$m_{0}$ is the bulk mass introduced as 
the non-anomaly mediated contribution 
in order to keep the squared slepton masses positive. 

In general, the contributions of the bulk mass are 
not universal among the particles, therefore, the coefficients 
$c_{Q}$, $c_{U}$, $c_{D}$, $c_{L}$, $c_{E}$, $c_{h_{u}}$, 
and $c_{h_{d}}$ can be different from one another. 
However, in the {\em minimal} AMSB (mAMSB) model \cite{Randall}, 
they are universal and normalized as: 

\begin{equation}
c_{Q}=c_{U}=c_{D}=c_{L}=c_{E}=c_{h_{u}}=c_{h_{d}}=1 \ . 
\label{eq:cmin}
\end{equation}

The mass relations, Eq.(\ref{eq:m2q}) to (\ref{eq:m2hd}), are 
Renormalization Group (RG) invariant except 
the bulk mass terms proportional to $m_{0}$. 
Since the smaller Yukawa couplings are neglected, 
the bulk mass terms are running only in the third generation. 
When $\tan{\beta}$ is enough small, 
only the top Yukawa coupling is effective. 
However, if $\tan{\beta}$ is large, 
not only the top but also the bottom Yukawa coupling is influential. 
Thus, we put both of these two contributions into the RG part of 
the third generation of the mass relations \cite{Huitu}: 

\begin{equation}
\frac{d}{dt} \left(
\begin{array}{c}
\delta m^{2}_{h_{u}} 
\\ \delta m^{2}_{U_{3}} 
\\ \delta m^{2}_{Q_{3}} 
\\ \delta m^{2}_{D_{3}} 
\\ \delta m^{2}_{h_{d}}
\end{array}
\right)
=
\frac{\alpha_{t}}{2\pi}
\left( \begin{array}{c}
3 \\ 2 \\ 1 \\ 0 \\ 0
\end{array}
\right)
\Delta m^{2}_{t}
+\frac{\alpha_{b}}{2\pi}
\left( \begin{array}{c}
0 \\ 0 \\ 1 \\ 2 \\ 3
\end{array}
\right)
\Delta m^{2}_{b} \ , 
\label{eq:rgdm2}
\end{equation}

with 

\begin{equation}
\Delta m^{2}_{t}
=
\delta m^{2}_{h_{u}} + \delta m^{2}_{U_{3}} + \delta m^{2}_{Q_{3}}
\ , 
\label{eq:dmt2}
\end{equation}

\begin{equation}
\Delta m^{2}_{b}
=
\delta m^{2}_{h_{d}}+\delta m^{2}_{D_{3}}+\delta m^{2}_{Q_{3}}
\ , 
\label{eq:dmb2}
\end{equation}
\noindent
where, $\delta m^{2}$'s are RGE running parts of the bulk mass terms. 
The initial values of these parameters are obviously 
($\delta m^{2}_{h_{u}}$, 
$\delta m^{2}_{U_{3}}$, 
$\delta m^{2}_{Q_{3}}$, 
$\delta m^{2}_{D_{3}}$, 
$\delta m^{2}_{h_{d}}$)
=
($c_{h_{u}}m^{2}_{0}$, 
$c_{U}m^{2}_{0}$, 
$c_{Q}m^{2}_{0}$, 
$c_{D}m^{2}_{0}$, 
$c_{h_{d}}m^{2}_{0}$). 
We evaluate these contributions numerically. In our framework, 
$m_{0}$ is introduced at the GUT scale of the gauge coupling 
constants $M_{X} \sim 2 \times 10^{16}$ (GeV) \cite{Arason}. 
The RG-variable $t$ is dimensionless and defined as 
$t=\ln(\sqrt{s}/M_{X})$ with the energy scale $\sqrt{s}$. 
Additionally, in the gaugino sector, we take into account 
the next-to-leading order radiative corrections and 
the weak scale threshold correction \cite{Gherghtta}. 

\section{Numerical Analyses} 

By making use of the routine \textit{FeynHiggs} \cite{HEINEMEYER}, 
we calculate the Higgs mass up to the full 2-loop level. 
As the corrections to the Higgs potential, up to 1-loop order 
of the top and bottom corrections \cite{Boer} are taken into account. 
The Higgs mixing parameter $\mu$ and 
the bilinear coupling constant B 
of the Higgs potential are given by following equations: 

\begin{equation}
|\mu|^{2}=\frac{m^{2}_{h_{d}}-m^{2}_{h_{u}}\tan^{2}{\beta}-
\frac{1}{2}m^{2}_{Z}(\tan^{2}{\beta}-1)+\Delta^{(1)}_{1-loop}}
{\tan^{2}{\beta}-1+\Delta^{(2)}_{1-loop}}
>0
\ , 
\label{eq:myu2}
\end{equation}

\begin{equation}
B=\frac{(m^{2}_{h_{d}}+m^{2}_{h_{u}}+2|\mu|^{2})\sin{2\beta}}
{2|\mu|\mbox{sgn}(\mu)}+\Delta^{(3)}_{1-loop}
\ , 
\label{eq:B}
\end{equation}
\noindent
where, $m_{h_{u}}$ and $m_{h_{d}}$ are the Higgs mass parameters 
given by Eqs.(\ref{eq:m2hu}) and (\ref{eq:m2hd}), respectively. 
$ \Delta^{(i)}_{1-loop}$, (i=1,2, and 3) 
imply the 1-loop correction terms \cite{Boer}. 
The positivity of $|\mu|^{2}>0$ certifies the appropriate shape of 
the Higgs potential. sgn($\mu$), the sign of $\mu$ is 
fixed to positive (see subsection 3.1 ). 
Moreover, CP-odd Higgs mass is also corrected \cite{Boer} 
up to 1-loop order with top and bottom contributions. 
The theoretical value of the muon g-2 in AMSB models is evaluated 
up to 1-loop order both of the chargino and neutralino 
contributions \cite{Ibrahim1}. 
We assume that the CP-violating phase effect is negligible. 

\subsection{Parameter Space} 
In this subsection, the allowed parameter areas 
on the $m_{0}-m_{3/2}$ plane \cite{Chattopadhyay} 
are estimated in mAMSB model given by Eq.(\ref{eq:cmin}). 
They are shown in Figs. 1. 
Note that $a^{SUSY}_{\mu}$ is redefined as 
$a^{SUSY}_{\mu} \rightarrow a^{SUSY}_{\mu} \times 10^{-10}$ 
for convenience. 
We fixed sgn($\mu$) to positive, because we find no region on 
the parameter space with negative $\mu$ 
as far as $a^{SUSY}_{\mu}$ is positive. 
In Figs. 1, the left hand sides are excluded by the experimental 
limits of the squark and slepton masses \cite{Particle,Fusaoka}. 
Moreover, the lower regions are restricted from 
the Higgs potential condition given by Eq.(\ref{eq:myu2}), and 
the mass limits of the charginos and neutralinos \cite{Particle}. 
The contours of the Higgs mass are illustrated 
by the thin solid curves. 
The SUSY contributions of g-2 are drawn with the thick solid curves 
for $a_{\mu}^{SUSY}$=50, 40, 30, 20, and 10, respectively. 
Once reliable $a_{\mu}^{SUSY}$ like Eq.(\ref{eq:aepm}) is 
given by the experiments, 
only a small region of the corresponding narrow belt is 
allowed as the parameter space. 
In another words, the allowed parameter space on $m_{0}-m_{3/2}$ 
plane is strictly restricted from the experiments. 

\subsection{Upper bound of Higgs mass} 
The upper bound of the Higgs mass 
is evaluated in the allowed parameter space where is strictly 
restricted from the muon g-2 in this subsection. 
In the first step, we scan on the allowed parameter region 
constrained by the g-2 analysis in the previous subsection. 
In the next step, we find out the parameter point to 
maximize the Higgs mass bound on the allowed region. 
In the last step, we plot those maximal values of the Higgs mass 
bound as functions of $\tan{\beta}$ as shown in Fig. 2(a). 
Additionally, the lower bound is also obtained in the similar 
procedure. 
For instance, $\tan{\beta}=3$ gives $m_{h}>102$ (GeV), or 
$m_{h}>115$ (GeV) is derived when $\tan{\beta} \geq 10$ . 

\subsection{Other AMSB models} 
In this subsection, other possibilities rather than the minimal model 
Eq.(\ref{eq:cmin}) of the bulk mass contributions are considered. 
Since naive AMSB models have a common serious problem that 
the squares of the slepton masses are negative, 
several scenarios are proposed to recover this defect. 
One of them is the gaugino assisted AMSB model \cite{Kaplan}. 
This model contains the coefficients of the bulk mass as follows: 

\begin{equation}
c_{Q}=21/10,\ c_{U}=8/5,\ c_{D}=7/5,\
c_{L}=9/10,\ c_{E}=3/5,\
c_{h_{u}}=9/10,\ c_{h_{d}}=9/10 \ . 
\label{eq:cga}
\end{equation}

\noindent
Another proposal is the extra U(1) model \cite{Carena}, 
and this model suggests following conditions: 

\begin{equation}
c_{Q}=3,\ c_{U}=-1,\ c_{D}=-1,\
c_{L}=1,\ c_{E}=1,\
c_{h_{u}}=-2,\ c_{h_{d}}=-2 \ . 
\label{eq:cexu}
\end{equation}

\noindent
We also analyze these two models in the similar way 
to the previous subsection. 
Figure 2(b) shows the upper bounds of the Higgs mass 
in these models in the same procedure with the previous subsection. 
Nonetheless, we cannot find so large difference on the upper bounds 
among these models as shown in Fig. 2(b). 

More generally, the coefficients can be shifted freely from 1 
as far as the bulk mass terms certify the positivity 
of the squared scalar masses. 
In order to evaluate the effects on the Higgs mass bound 
of their free shifts from 1, we move one of 
$c_{Q},c_{U},c_{D},c_{L},c_{E},c_{h_{u}}$, or $c_{h_{d}}$ from 1 
(and all of the others are kept equal to 1 ). 
$\tan{\beta}=45$ and $a_{\mu}^{SUSY}$=10 are fixed, because 
this condition is the strictest to make the upper bound 
of the Higgs mass largest. 
As Fig. 3 shows, the coefficients except $c_{L}$ and $c_{E}$ 
give not so large contribution to the upper bound of the Higgs mass. 
Here is a note on the $c_{Q}$ and $c_{U}$. 
We found in our preliminary stage that their shifts 
appear to be sensitively increasing the upper bound 
if they are less than $-3$ . 
However, we also find their sensitivities 
are canceled with the conditions on the mass parameters 
of the other scalars not to acquire the VEVs 
as the charge and color breaking \cite{Ferreira}. 
Therefore, the upper bound of the Higgs mass is 
not so increasing by the free shift of $c_{Q}$ or $c_{U}$. 
The shifts of $c_{L}$ and $c_{E}$ from 1 contribute 
only to decrease the upper bound as shown in Fig. 3 . 
Note that $c_{L}$ and $c_{E}$ must be positive because 
these bulk mass terms are introduced to make 
the squared slepton masses positive. 

\ \\
\noindent
As the result, we conclude that $m_{h}<123$ (GeV) 
in almost AMSB models for all of $\tan{\beta}<$60 . 

\section{Summary and Conclusion} 
We estimate the upper bound of the Higgs mass 
in several anomaly mediated supersymmetry breaking models. 
Consequently, we find $m_{h}<123$ (GeV) in AMSB models. 
Heretofore, 
the upper bound of the Higgs mass has been estimated 
by loosely assuming some of the SUSY parameters. 
However, we achieve to confine the allowed parameter space 
into the restricted area on the $m_{0}-m_{3/2}$ plane, and 
estimate the upper bound without assuming the parameters 
except $\tan{\beta}$ by making use of 
the restrictions from recent muon g-2 measurements. 
Therefore, we obtain a reliable estimation on the upper bound. 
Needless to say, $m_{h}=123$ (GeV) as the upper bound of 
the Higgs mass is sufficiently small at present 
or in the next generation of the high-energy colliders. 
We hope to be found the Higgs boson in the near future. 

\section*{Acknowledgement}
J. Tabei would like to thank Y. Tanaka 
for useful advice and discussions.

%
%
%
%
\begin{figure}[htbp]
   \begin{center}
     \includegraphics[scale=0.3]{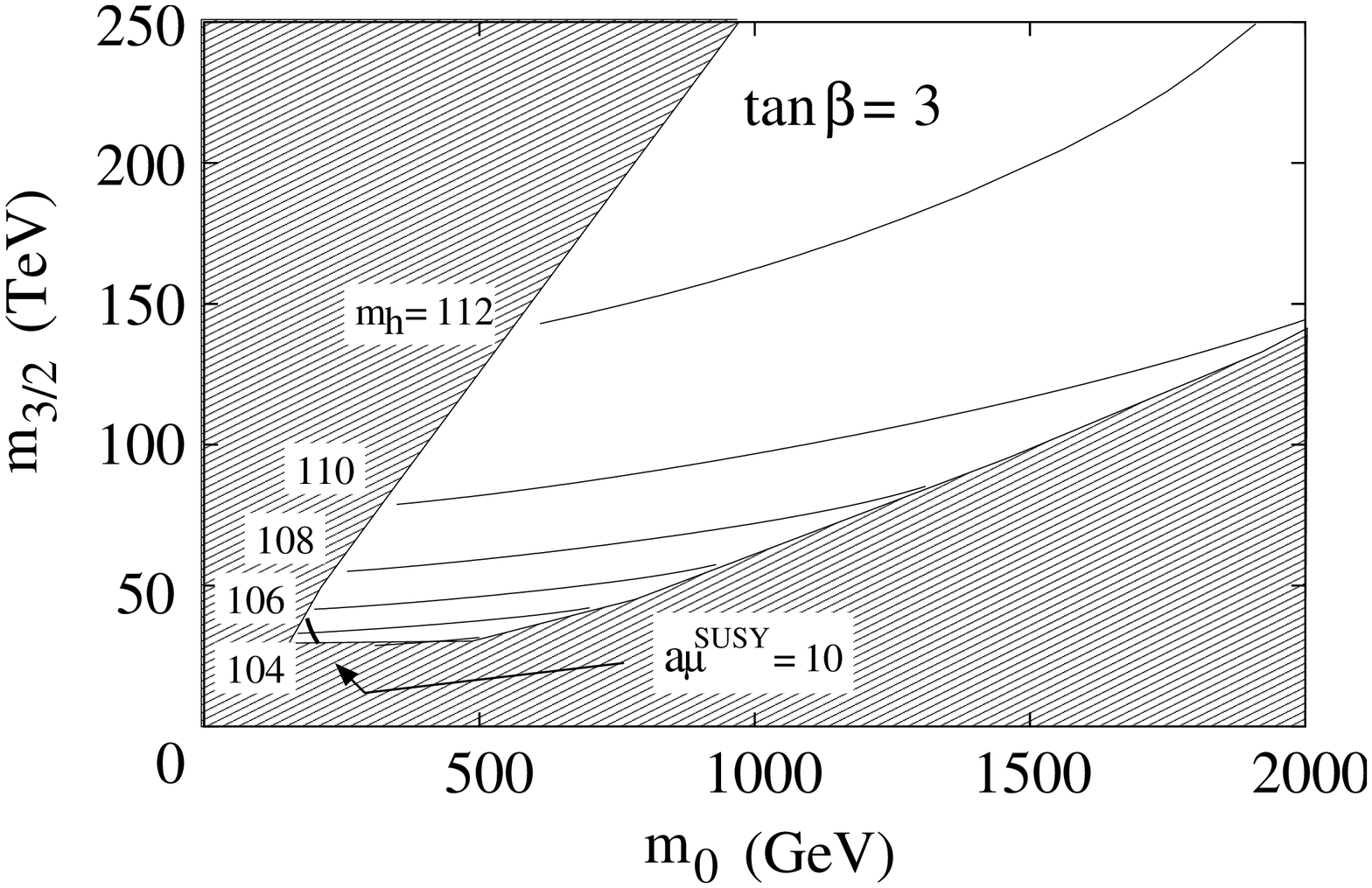}
     \includegraphics[scale=0.3]{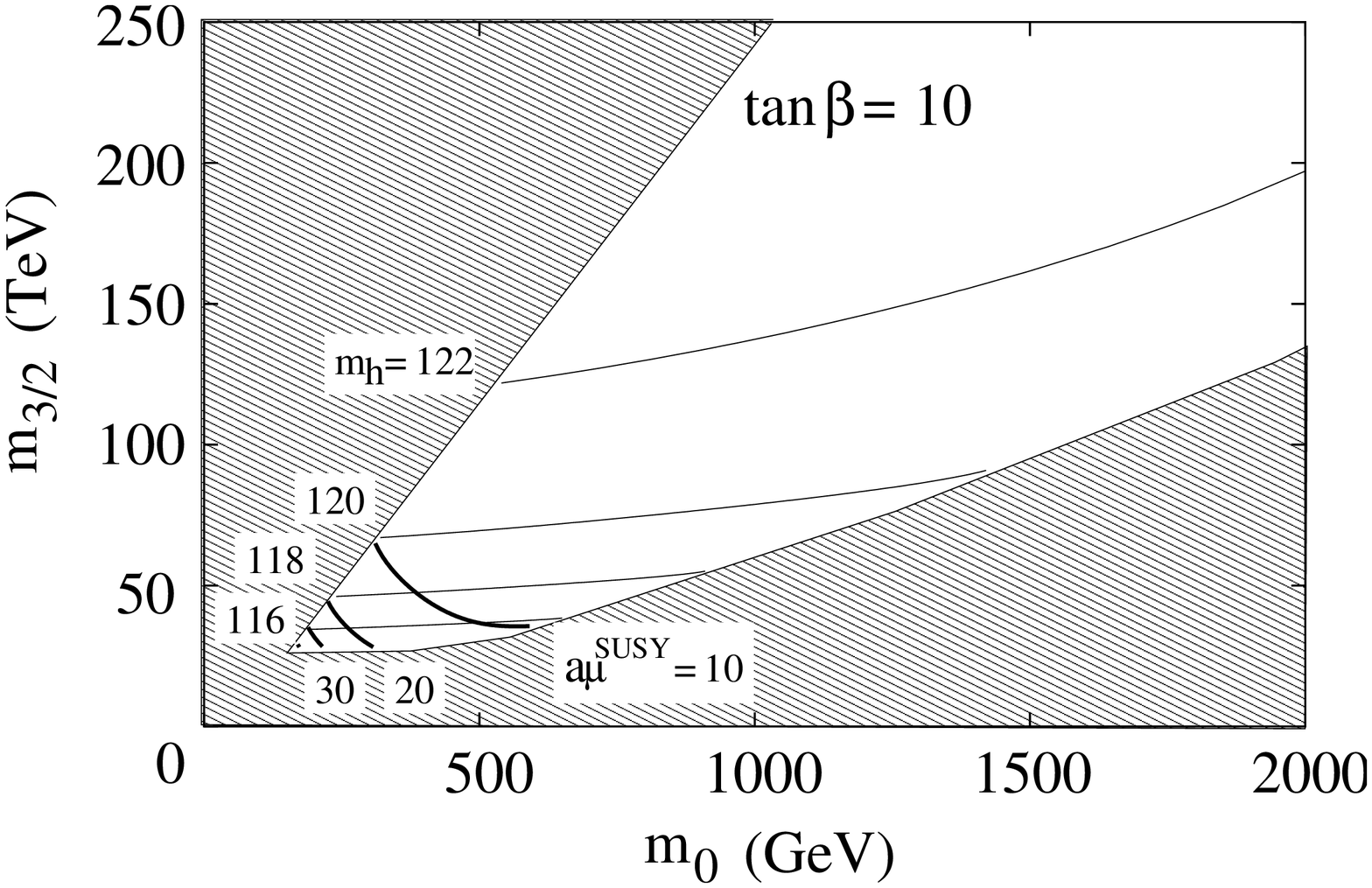}
     \includegraphics[scale=0.3]{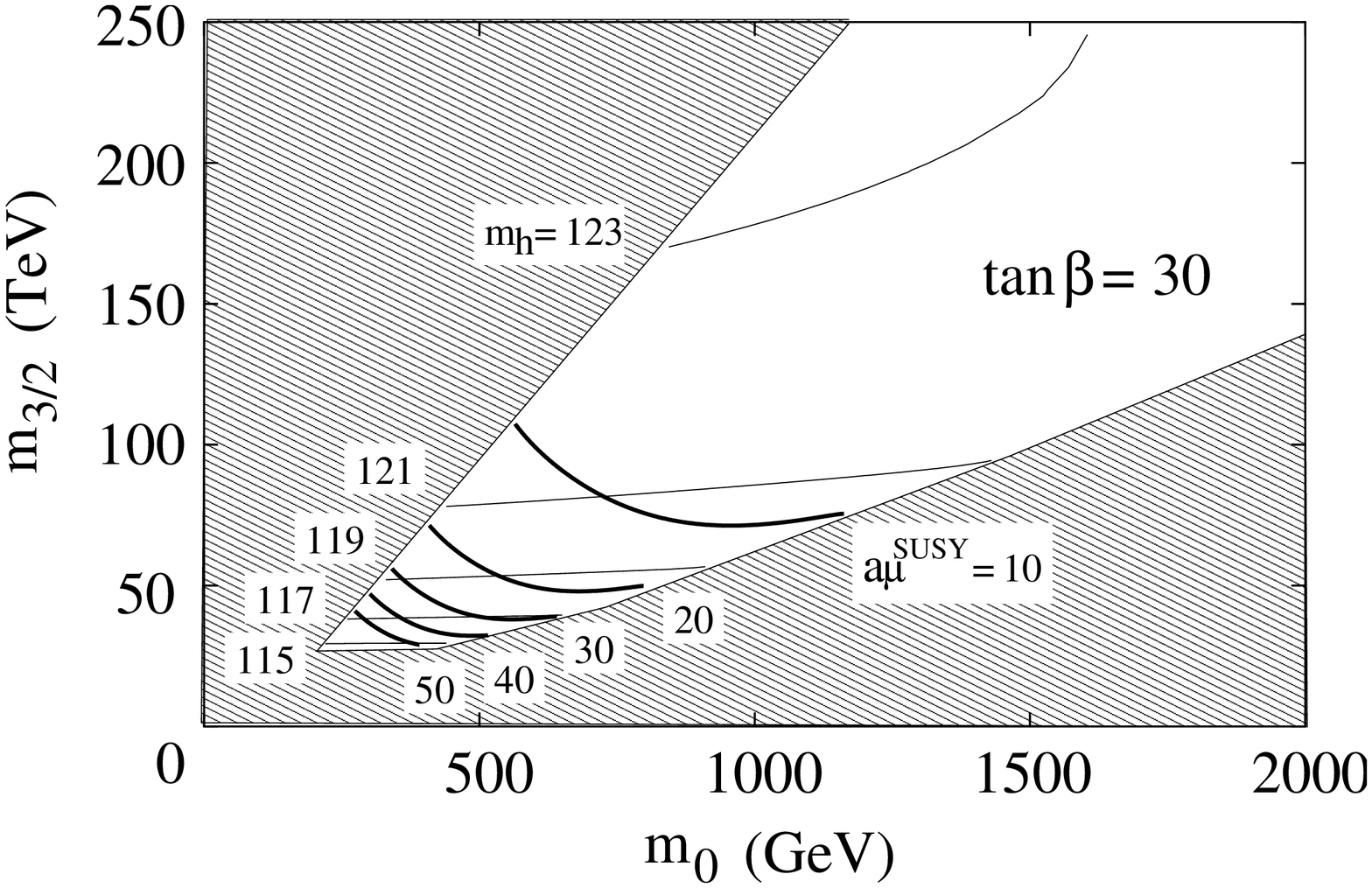}
     \includegraphics[scale=0.3]{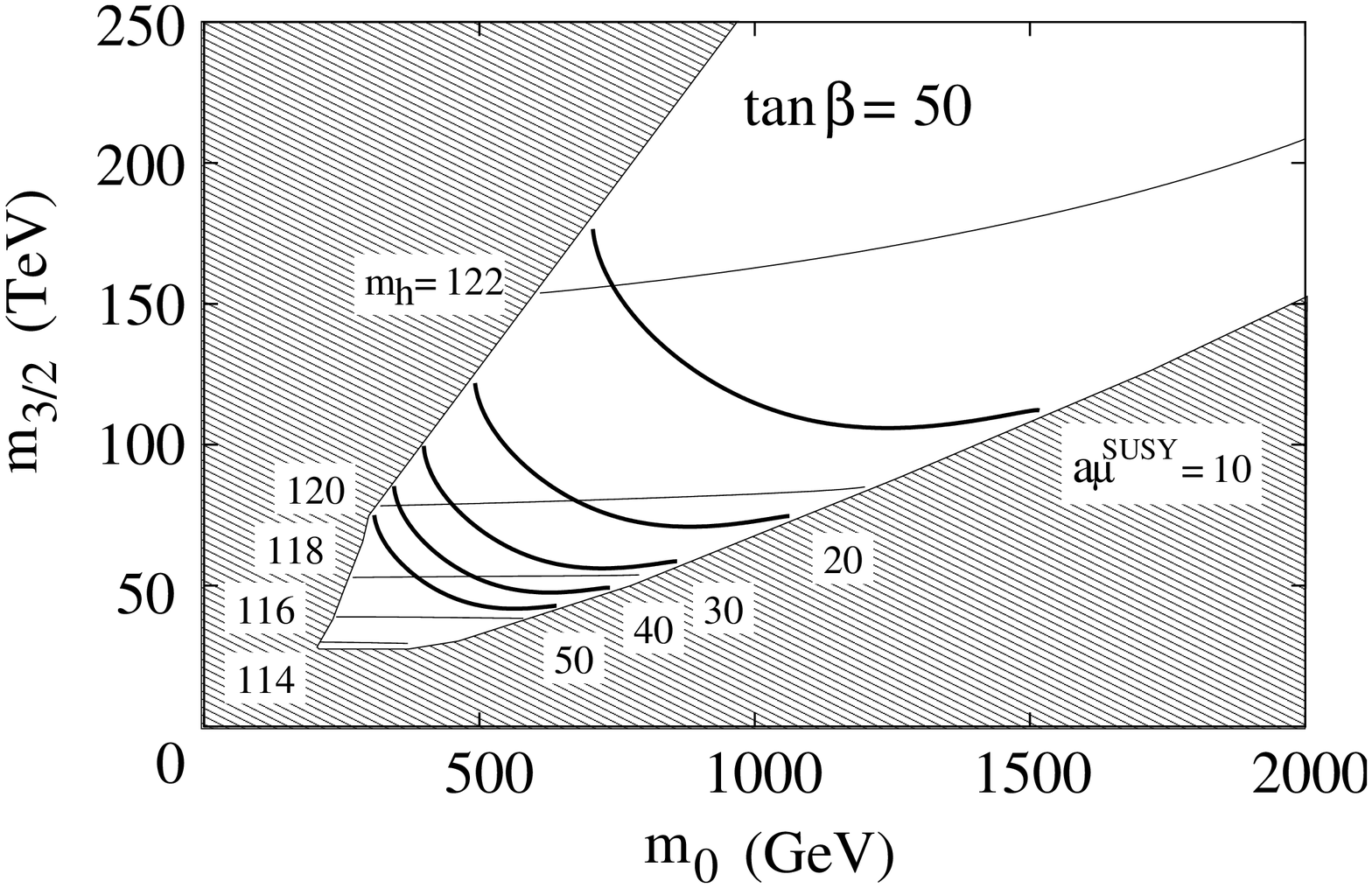}
\caption{$m_0-m_{3/2}$ planes with the Higgs mass (GeV) and 
$a^{SUSY}_{\mu}(\times 10^{-10})$ contours at several $\tan{\beta}$ 
in the mAMSB model. The upper left sides are excluded 
by the limits of the squarks and sleptons masses. 
Lower regions are excluded by the proper Higgs potential condition 
and the limits of the chargino and neutralino masses. }
   \end{center}
\end{figure}
\begin{figure}[htbp]
  \begin{center}
    \includegraphics[scale=0.3]{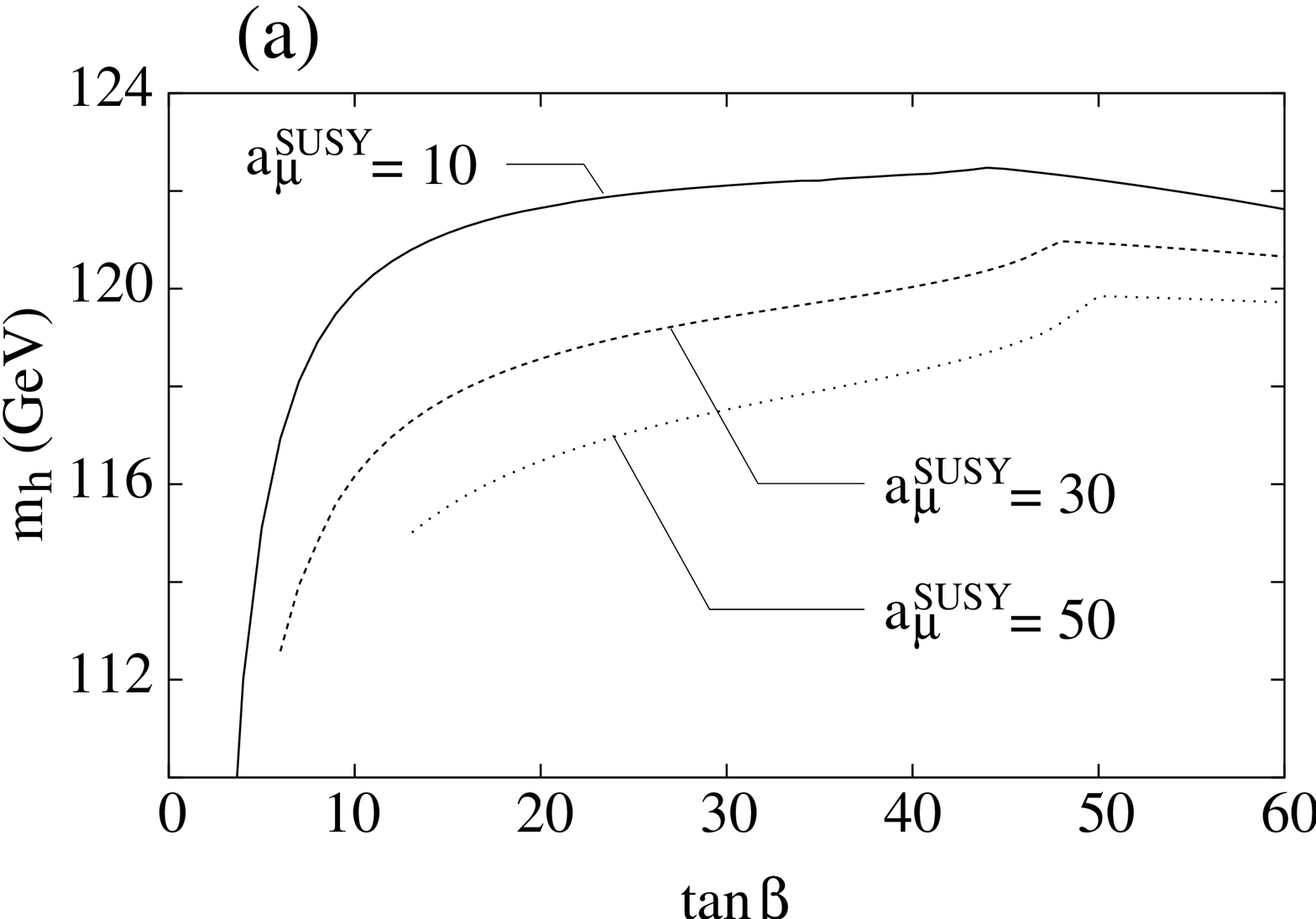}
    \includegraphics[scale=0.3]{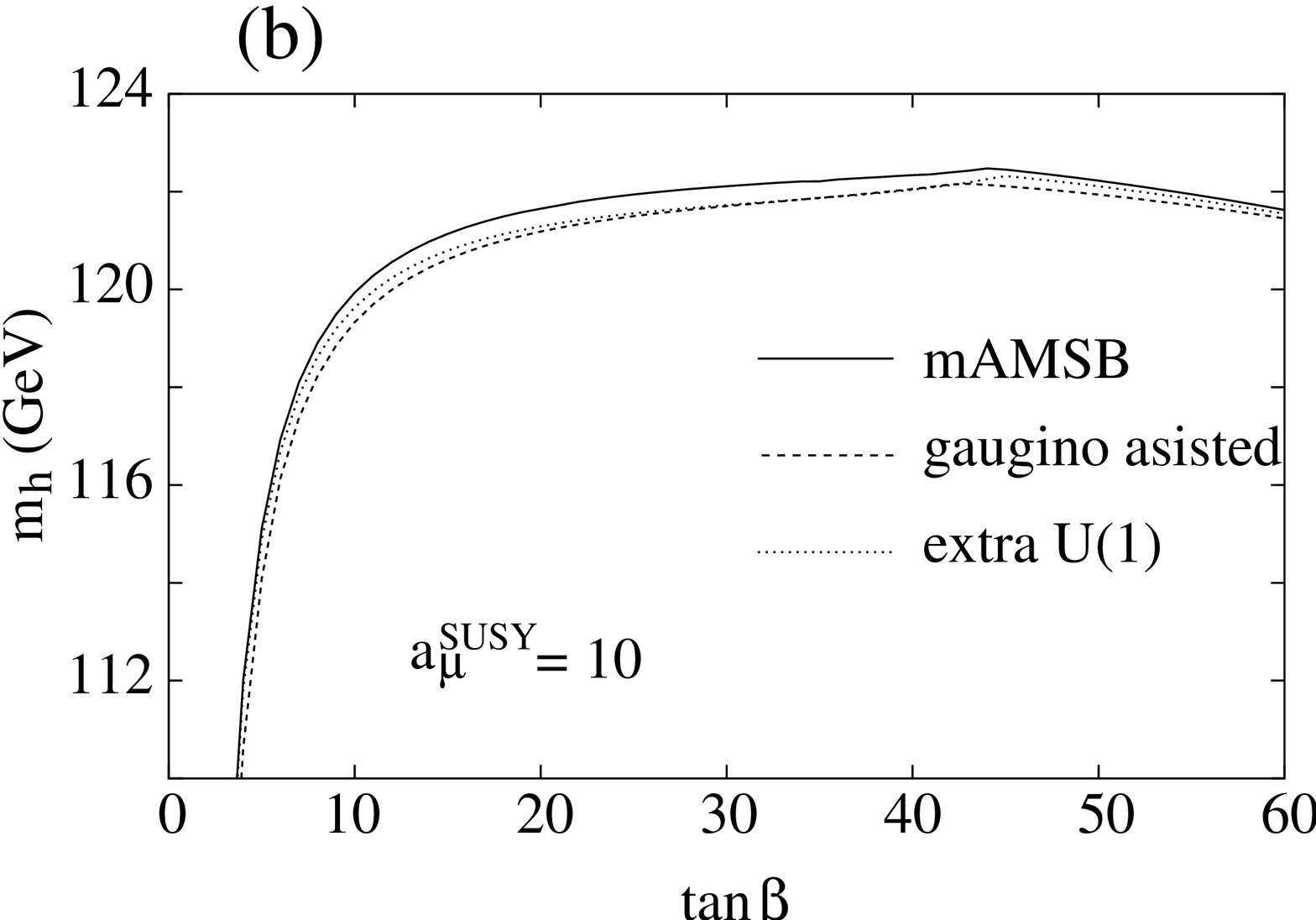}
    \caption{(a) The upper bounds of the Higgs mass depending on 
$\tan{\beta}$ for $a_{\mu}^{SUSY}$=10, 30, and 50 in the mAMSB. 
(b) The upper bounds of the Higgs mass 
as functions of $\tan{\beta}$ in three AMSB models. 
$a_{\mu}^{SUSY}$ is fixed to 10 . }
  \end{center}
\end{figure}
\begin{figure}[htbp]
  \begin{center}
    \includegraphics[scale=0.4]{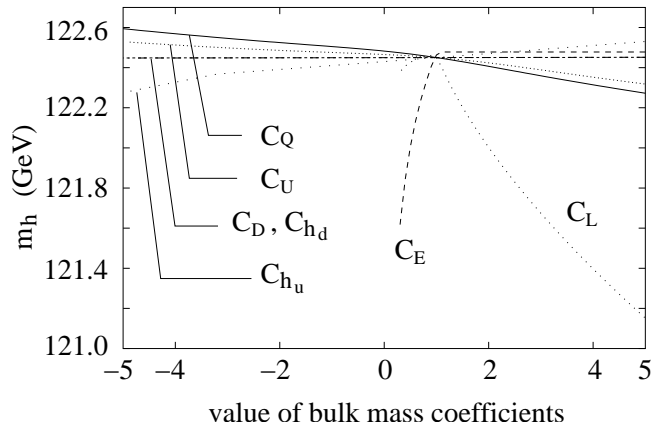}
    \caption{Effects on the Higgs mass bound from the bulk mass
coefficients. 
Only one of these seven coefficients is moved, and 
all of the others are fixed to 1 . 
$a_{\mu}^{SUSY}$=10 and $\tan{\beta}$=45 are fixed. }
  \end{center}
\end{figure}
\end{document}